\documentclass[twocolumn,prl,aps,groupedaddress,nopacs,footinbib]{revtex4}

\usepackage{amsmath}
\usepackage{amssymb}
\usepackage{latexsym}
\usepackage{graphicx}
\usepackage{hyperref}
\usepackage{mathrsfs}
\usepackage{color}
\usepackage{bm}

\newcommand{\be}{\begin{equation}}
\newcommand{\ee}{\end{equation}}
\def\bea{\begin{eqnarray}}
\def\eea{\end{eqnarray}}
\def\ba{\begin{eqnarray}}
\def\ea{\end{eqnarray}}

\begin{document}

\title{Direct reconstruction of dark energy}
\author{ Chris Clarkson$^1$ and Caroline Zunckel$^{2,3}$   \\
\it $^1$Centre for Astrophysics, Cosmology \& Gravitation, 
and, Department of Mathematics and Applied Mathematics, 
University of Cape Town,
South Africa 
\\
\it $^2$Astrophysics Department, Princeton University, 
New Jersey, USA \\
\it $^3$Astrophysics and Cosmology Research Unit, 
University of KwaZulu-Natal, 
South Africa
}

\begin{abstract}

An important issue in cosmology is reconstructing the effective dark energy equation of state directly from observations. With so few physically motivated models, future dark energy studies cannot only be based on constraining a dark energy parameter space.  We present a new non-parametric method which can accurately reconstruct a wide variety of dark energy behaviour with no prior assumptions about it. It is simple, quick and relatively accurate, and involves no expensive explorations of parameter space. The technique uses principal component analysis and a combination of information criteria to identify real features in the data, and tailors the fitting functions to pick up trends and smooth over noise. We find that we can constrain a large variety of $w(z)$ models to within 10-20$\%$ at redshifts $z\lesssim1$ using just SNAP-quality data. 
\end{abstract}

\maketitle

\paragraph{Introduction} 
The dark energy crisis in cosmology highlights our incomprehension of what the universe actually consists of. Usually characterised by an effective equation of state function $w$ in which we hide our lack of understanding, an important goal over the coming years will be to try to understand this variable as a function of redshift, giving $w(z)$. Theoretically we have little to go on, other than the cosmological constant which has $w=-1$ for all time. Alternatives such as quintessence or modified gravity theories make predictions about how $w$ diverges from $-1$ yet are often forced to parameterise free functions \cite{copeland}. More radically, proposals which modify the radial distribution of matter on Hubble scales provide no a priori constrains on what the effective equation of state (defined by matching up the distance indicator with an FLRW model) could be \cite{FRLW}. So, the forward problem of parameterising models, matching to data and discarding the fits which are poor or over-parameterised, suffers from a profound arbitrariness when we try to interpret the errors:  using simple smoothly varying functions of $z$ severely limit the departures from cosmological constant to a small range of models, but if we add more freedom to $w$ the errors grow uncontrollably. Without encapsulating the behaviour of $w(z)$ which may actually exist, what are we really constraining? Would a `backwards' method be better? Can we instead reconstruct $w$ from observations directly?

The dark energy equation of state is typically (re)constructed using distance measurements as a function of redshift. 
The luminosity distance may be written as 
$d_{L}(z)=\frac{c(1+z)}{H_0 \sqrt{-\Omega_k}}\sin{\left( 
\sqrt{-\Omega_k}\int_0^z{\mathrm{d}z'\frac{H_0}{H(z')}}\right)},
$ where $H(z)$ is given by the Friedmann equation, 
$H(z)^2= H_0^2\{\Omega_{m} (1+z)^3+\Omega_{
k}(1+z)^2
+(1-\Omega_m-\Omega_k)\exp{[3\int_0^z
\frac{1+w(z')}{1+z'}\mathrm{d}z']}\},
$ where $H_0=H(0)$ and $\Omega_{m,k}$ are the normalised density parameters.  A common procedure is to postulate a several parameter form for $w(z)$ and calculate $d_L(z)$. The most promising of these approaches uses a principal component analysis to construct the `optimal' basis functions for $w(z)$ based on the data~\cite{HS}.
An alternative method is to reconstruct $w(z)$ by directly reconstructing the luminosity-distance curve.  Writing $D(z)=(H_0/c)(1+z)^{-1}d_L(z)$ as the normalised comoving distance, we have~\cite{Starobinsky:1998fr,Nakamura:1998mt,Huterer:1998qv}: 
\ba
&&\!\!\!w(z)=\{2(1+z)(1+\Omega_kD^2)D''-[(1+z)^2\Omega_kD'^2\nonumber\\&&
+2(1+z)\Omega_kDD'-3(1+\Omega_kD^2)]D'\}/\\&&
\{3\{(1+z)^2[\Omega_k+(1+z)\Omega_m]D'^2-(1+\Omega_kD^2)\}D'\}.\nonumber
\label{w}
\ea
Thus, given a distance-redshift curve $D(z)$, we can reconstruct the dark energy equation of state, assuming we know the density parameters $\Omega_m$ and $\Omega_k$. Different methods for doing this involve smoothing the data to give $D(z)$, or parameterising $D(z)$ by a function; see~\cite{Sahni:2006pa} for a comprehensive review, and~\cite{weller_albrecht,Alam:2003sc,daly,Alam:2004jy,Wang:2004py,Daly:2004gf,shafieloo} for alternative model independent approaches.

The direct reconstruction method is unstable because of the two derivatives of the observed function in eq. \ref{w}, requiring the fitting function to accurately capture the slope and concavity of luminosity distance curve.  This means that differences between the true underlying model and the fitted function due to the choice of parameterisation, are amplified drastically when reconstructing $w$. Furthermore, $w$ is constructed from a quotient of functions which need to balance to obtain the correct $w$. However, there must exist a set of `optimal' basis functions with which achieving this balance becomes possible and direct reconstruction feasible; in essence this is the same as specifying `the correct' parameterisation for dark energy. Can we find such a basis?

\begin{figure*}[htbp]
\begin{center}
\includegraphics[width=0.34\textwidth]{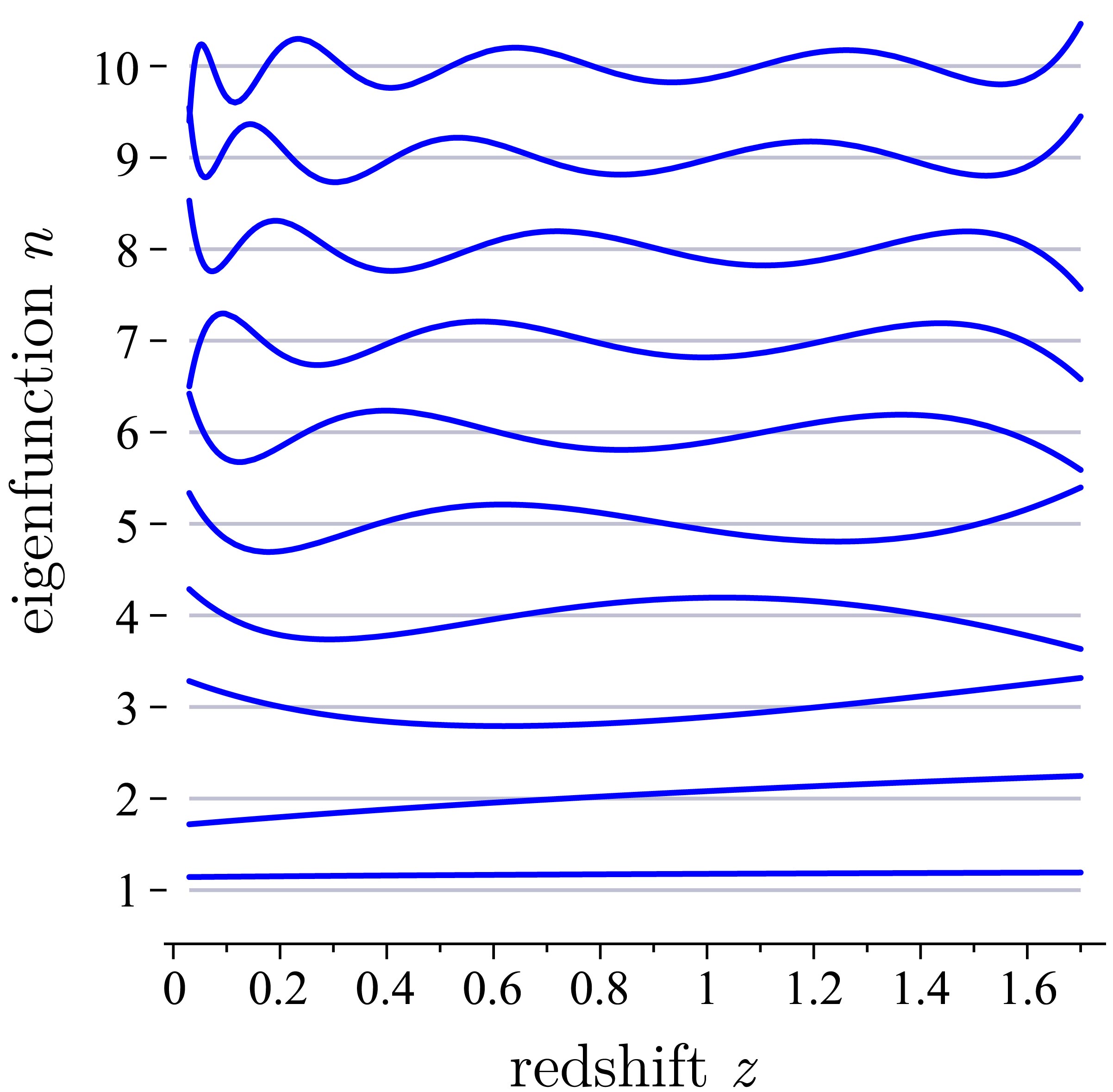}
\includegraphics[width=.64\textwidth]{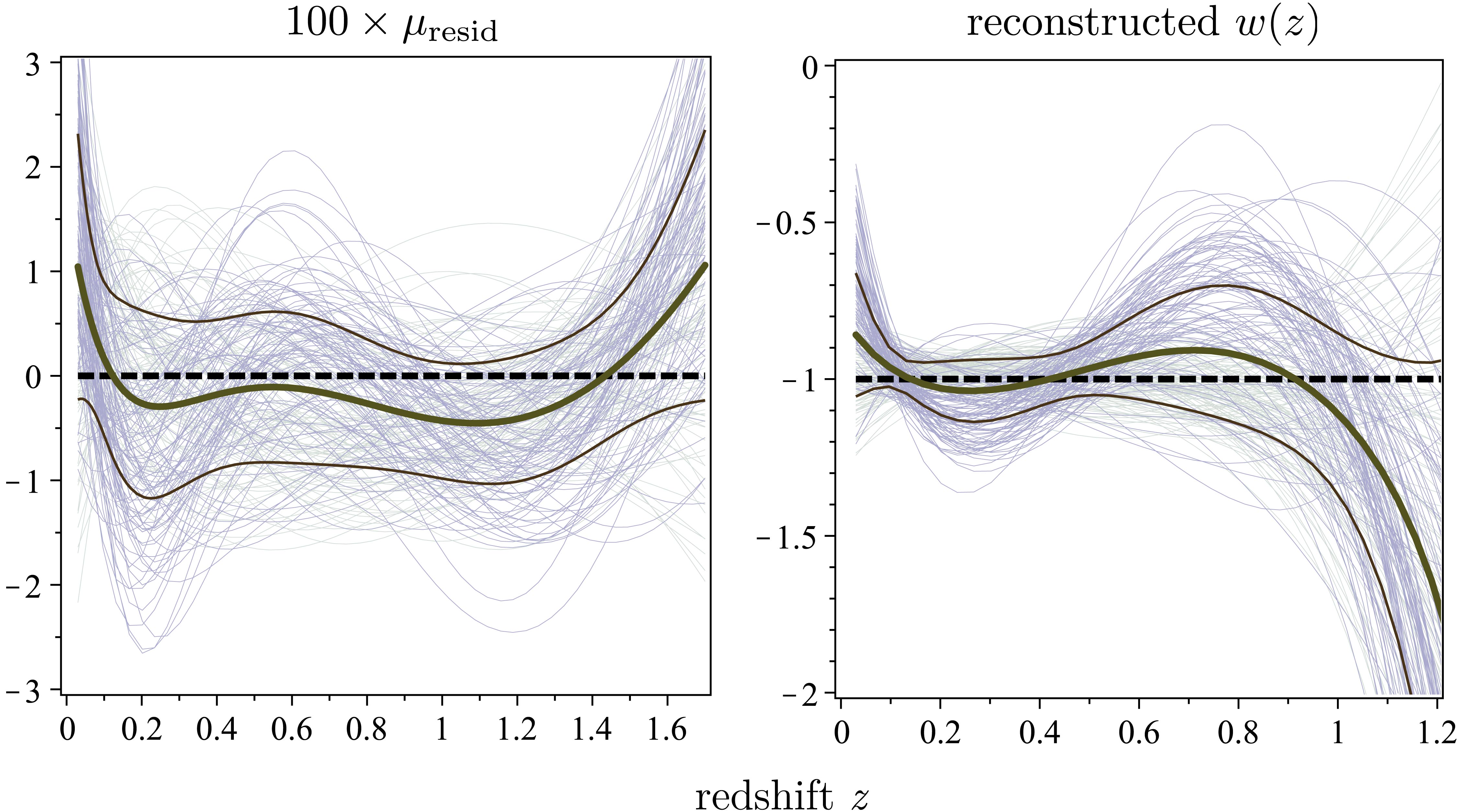}
\caption{Extracting $w=-1$. The first 10 eigenfunctions for $\Lambda$ are shown left. We take two separate reconstructions, using $M=4$ and $M=5$ ($N=10$), and generate the errors on $\mu_\text{resid}$ via a monte carlo (with 100 samples each for clarity). The two reconstructions are combined with equal weight to give the 1-$\sigma$ errors. The choice of number of eigenfunctions is not yet favourable; beyond $z\sim1.2$ (not shown) the reconstruction is hopeless for these eigenfunctions. }
\label{recon}
\end{center}
\end{figure*}

At first sight it seems not: If we have no inherent intuition of the true $w$, how can we possibly guess the right form for $D(z)$? Of course, a large polynomial expansion would work but at the expense of ludicrous errors. Furthermore, one can easily achieve a fit that is too good: a $\chi^2$ is less than the value the actual underlying model would produce. Over-fitting to noisy data translates into wild behaviour in $w$. 

Here we present a method to find well adapted basis functions for fitting $D(z)$ and a simple way to construct the errors on $w$. We first calculate the residuals of the measured apparent magnitude around some fiducial model. Assuming some basis functions,  we then calculate the principal components of the residuals, and use those fixed principal components to provide a measure of the true $D(z)$.
Given the many possibilities, a combination of information criteria are used to select which encompass the information present in the data. Folding the errors together appropriately produces a non-parametric method which reproduces $w(z)$ together with an effective `1-$\sigma$' confidence measure. A non-parametric method doesn't produce confidence limits; rather a confidence interval of $x$\% must be expected to trap the correct value $x$\% of the time~\cite{non-param}.

\paragraph{Basis functions for distances}

Let's assume we have $N_d$ data points distributed at $z_i$ for the distance modulus $\mu=5\log_{10}d_L(z)+25$, with Gaussian errors $\sigma_i$ which are independent. 
We create $\mu_\text{resid}=\mu_\text{data}-\mu_\text{fiducial}$ where the fiducial model is some predefined model, such as flat LCDM, an empty model, or EdS. A good choice is the best-fit LCDM model, being consistent with current data. Our goal is to construct $\mu_\text{resid}$ and two derivatives as accurately as possible.  We choose a set of primary basis functions $p_n(z)$ such as $p_n(z)=z^{n-1}$ and use  the function
$\sum_{n=1}^{N} a_n p_n(z)
$ as the basis to fit $\mu_\text{resid}(z)$ to data, for a fixed $N$. This is a linear fit, so can be done easily, and the covariance matrix $\bm C$ calculated algorithmically without having to explore a complicated parameter space, or make assumptions about Gaussian errors on the parameters. Now we perform a principal component analysis on the fit to find the best basis of functions as follows.  We diagonalise the inverse covariance matrix, and create a matrix of eigenvectors $\bm E$. We then order them according to decreasing eigenvalue; i.e., so that the $n$'th column of $\bm E$, $\bm e_{n}$, is the eigenvector with the $n$'th largest eigenvalue, etc. Write $e_{mn}$ as the $m$'th component of the $n$'th eigenvector. 
 Define the eigenfunctions
$
P_n(z)=\sum_{m=1}^N e_{mn} p_m(z),\text{~for~}n\in[1,N]
$ 
which now form a new family of basis functions which are suitably adapted to the data. In matrix form $\bm P=\bm E^T\bm p \Leftrightarrow \bm p=\bm E\bm P$. These functions are now orthogonal with respect to the errors on the data:
$
\sum_i^{N_d} \sigma_i^{-2}P_m(z_i)P_n(z_i)=0\text{~if~} m\neq n.\,
$
In other words, the Fisher matrix in this basis is diagonal. We can normalise the eigenfunctions
$\hat P_n(z)={P_n(z)}/{\sqrt{\sum_i^{N_d} \sigma_i^{-2} P_n(z_i)^2}},$ in which case the Fisher matrix will be the identity matrix. If we now refit to the data with new parameters for these basis functions, the covariance matrix is also (very nearly) the identity matrix. 
For a fixed $N$, we may be interested in the first $M<N$ eigenfunctions which encompass the dominant features in the data, and throw away the higher ones which contain noise-induced oscillations.

\begin{figure*}[htbp]
\begin{center}
\includegraphics[width=0.7\textwidth]{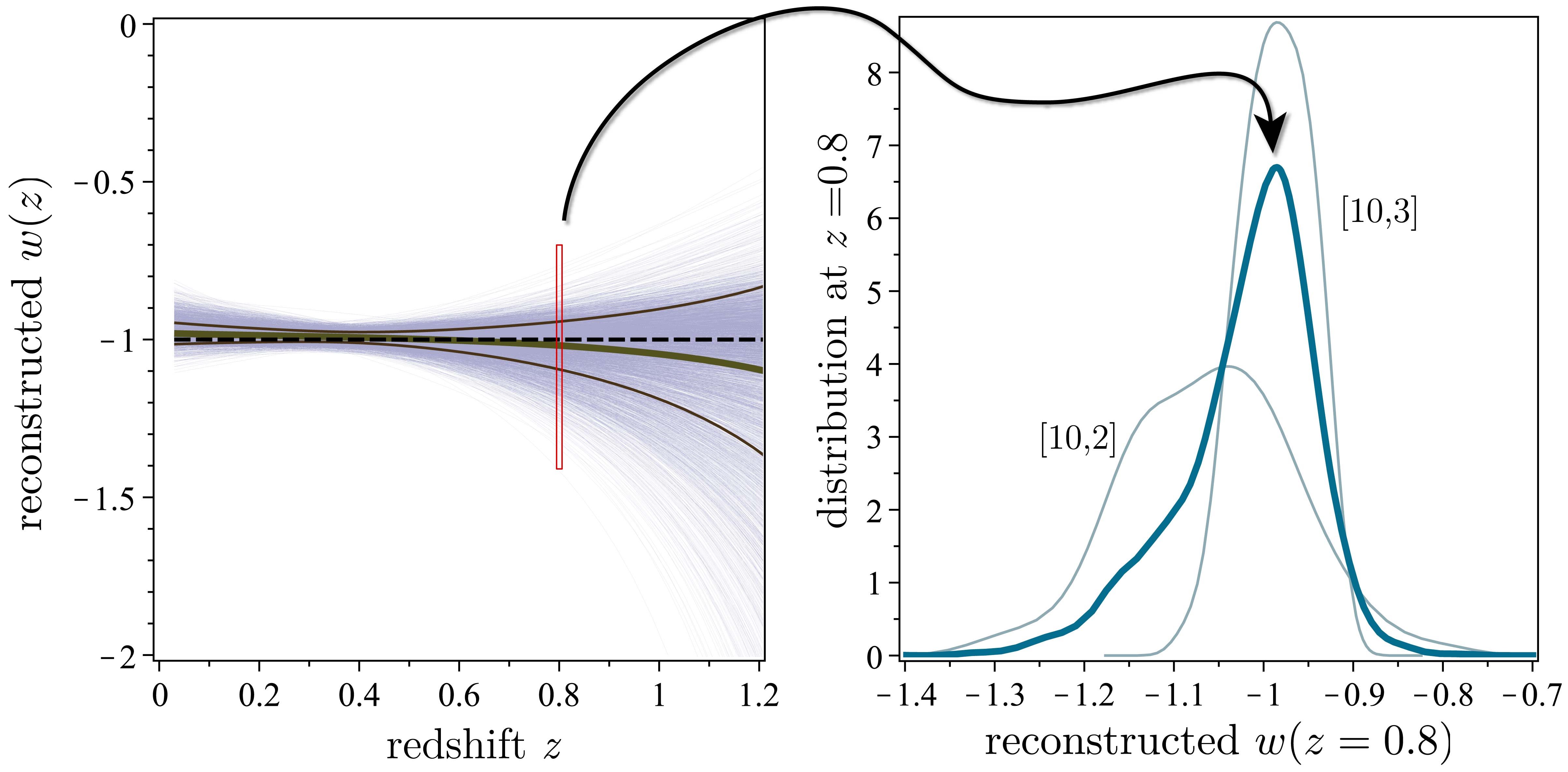}\includegraphics[width=0.3\textwidth]{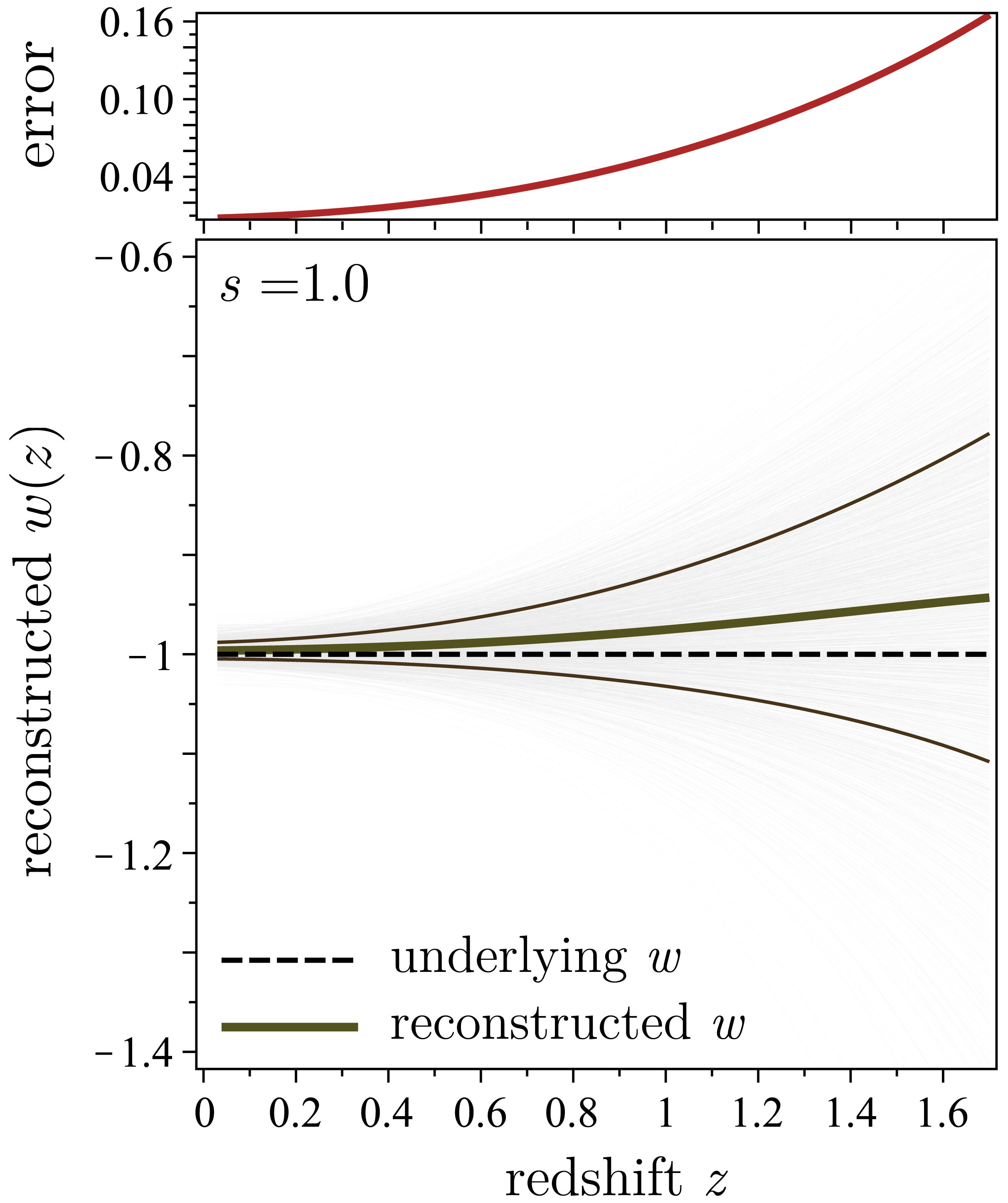}
\caption{Reconstruction of $w=-1$ using the CIC criteria, Eq.~(\ref{cic}), with $s=0.2$. All $[N,M]$ values satisfying Eq.~(\ref{cic}) are monte carlo-ed using 200 runs each (left); the results are bundled into a single probability distribution (middle) from which we infer 1-$\sigma$ errors at each redshift, shown left. For this example, the CIC selects for $M=2,3$ for each $N$. We show two probability distributions for $N=10$ (middle), as well as the combination for all $N,M$ used.  It is at the stage of bundling up the different $N,M$ eigenmodes into one probability distribution where the method becomes non-parametric. The combination of different eigenmodes leads to slight non-Gaussianity of the distribution. In this example, $s$ can be increased all the way up to 1 which shrinks the errors even further, and allows tight constraints on $w$ over the full range of $z$ (right).}
\label{lambda}
\end{center}
\end{figure*}

\paragraph{Practicalities and a test case - can we recover $\Lambda$?}
To test the method, we construct a hypothetical data set in line with expectations from the SNAP experiment, consisting of 2000 type 1a supernovae (SNIa) measurements evenly distributed in the redshift range $z=0.08-1.7$, and 300 local SNIa in $0.03-0.08 $~\cite{SNAP}. The statistical uncertainty  is conservatively estimated as $\sigma_{mag}=0.15$mag and we include systematic errors, as a linear drift from $\sigma_{sys}=0-0.02$ \cite{SNAP}.  We assume a true underlying $w$ and some parameters, $\zeta^\text{true}=\{ \Omega_m,\Omega_k,h\}^\text{true}=\{0.3,0.0,0.65\}$.  Our goal is to recover $w_\text{true}(z)$ in a robust way. We assume $\zeta^\text{fiducial}=\zeta^\text{true}$ for now, and take $w_\text{fiducial}=-1$. The actual model used for $\mu_\text{fiducial}$ is re-incorporated when evaluating $w(z)$, but an incorrect choice of the parameters for $\zeta^\text{fiducial}$ are subsumed by the usual uncertainties on those parameters. If we pick a fiducial model which is reasonably close to the true underlying cosmology, the residuals are predominantly noise and variations in $w$, requiring fewer fitting parameters giving typically smaller errors. Errors from the incorrect choice of $\zeta$ will be considered elsewhere but are standard. Here, we consider only the reconstruction errors for clarity.

  We used a variety of primary basis functions such as $p_n(z)\in \left\{z^{n-1},\left[{z}/(1+z)\right]^{n-1},\left[{1}/(1+z)\right]^{n-1}\right\}$, with similar results, though we achieve smoother reconstructions as we move left to right in this list; we present our results using the middle one. For a fixed $N$ the eigenfunctions range from smooth with no turning points, to very oscillatory; typically, the $n$'th eigenmode crosses zero $n-1$ times. Roundoff error can cause problems for $N\gtrsim10$, which is signalled by $\bm C$ having non-zero diagonal elements for the normalised basis. (Strictly speaking, $\bm C$ differs from the identity matrix slightly, since we use one realisation of the data: we would expect a change of roughly $1/\sqrt{N_d}$ for the diagonal elements.) For $N>10$ a singular valued decomposition could be used for the fits. In Fig.~\ref{recon} we show the first 10 normalised eigenfunctions when the underlying model is $w=-1$, with $N=10$. 
The first $M$ of these eigenmodes are used as the new basis functions for $\mu_\text{resid}$, and these are fitted to the data.  In the normalised basis the errors on the parameters are all Gaussian and unity (up to $\sim1/\sqrt{N_d}$).  Since the parameters are uncorrelated, the errors may be propagated into $\mu_\text{resid}$ by a simple monte carlo for each parameter. These curves then give a family of $w(z)$ curves from which the 1-$\sigma$ error may be given.  Errors on $\zeta$ may be folded in at this stage and lead to larger error bars, though we don't investigate this here.
We show an example of this reconstruction procedure in Fig.~\ref{recon}, using $M=4$ and 5, mixed together to form one set of error bars. There is nothing to say that these are good choices of $M$, an issue we explore now.

\begin{figure*}[t]
\begin{center}
\includegraphics[width=0.8\textwidth]{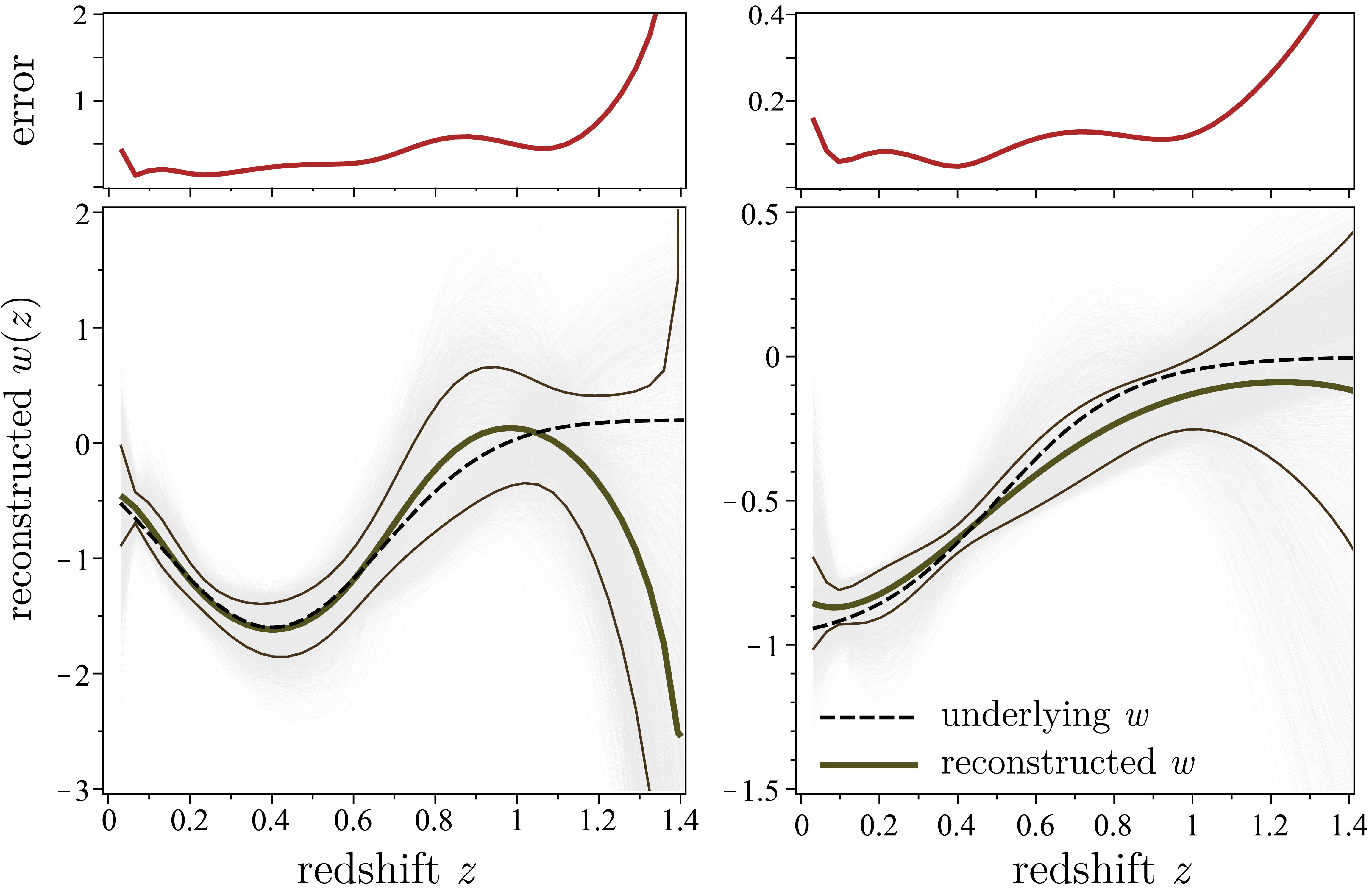}
\caption{Reconstruction of evolving dark energy. With a choice of $s=0.2$ and $\kappa=5$ we can reconstruct $w$ to $z\sim1$ with $\sim10$-20\% accuracy.}
\label{ws}
\end{center}
\end{figure*}

\paragraph{Selection criteria}
The number of eigenfunctions to use in the final reconstruction is critical as it determines the accuracy and size of the errors bars. Consider a subset of $[N,M]$ with $N\in[2,10]$ and $M\in[2,N]$.  Each choice $[N,M]$ will give a particular reconstruction of $w(z)$, together with some errors.  In the case where $M=N$, the original error bars are recovered and no information is thrown away. Reducing $M$ is accompanied by a reduction in the error, but an increased chance of getting $w(z)$ wrong. How do we select the `correct' set of eigenfunctions? Choosing the combination $[N,M]$ leading to the smallest $\chi^2$ runs the risk of overfitting to noise.  The Risk may be used~\cite{non-param,HS}, but requires the knowledge of the underlying value of $w(z)$ \emph{a priori}. 
 
Instead, we use a mixture of Akaike and  Bayesian information criteria~\cite{IC}:
\be
\text{AIC}=\chi^2_\text{min}+2M
,~~~\text{BIC}=\chi^2_\text{min}+M\ln N_d\, ,
\ee
where smaller values are assumed to imply a more favoured model.  The utility of these criteria over the Risk is that they are computed without knowing the underlying solution. We evaluated AIC and BIC values corresponding to each $[N,M]$ combination for a number of test cases and found that minimizing these two criteria lead to dramatically different reconstructions that were usually not optimal. Typically, models with a very low BIC are too smooth with tight error bars, while those with low AIC values are often too oscillatory and have large errors.  A more adaptable requirement uses a combined information criteria which we define as:
\be
\text{CIC}=(1-s)\, \text{AIC}+s\, \text{BIC},
\ee 
where $s$ takes us from conservative models when $s=1$ to more wild models when $s=0$.  The parameter $s$ thus mediates the competition between a smoother reconstruction (in which BIC is minimized) and one that is more featured (in which the AIC criterion is smallest). 

But there is no reason to select one particular reconstruction; the minimum CIC is still no silver bullet. We find a successful strategy is to select different $[N,M]$ choices which are near the best values of CIC, for a given $s$, and amalgamate them at the monte carlo stage when we compute the errors. We weight each $[N,M]$ choice equally. In this way, we reduce inherent bias which exists in any \emph{particular} choice of $[N,M]$, even after the principle component analysis.

After experimenting number of different $w(z)$, we find that the family of $[N,M]$ reconstructions which satisfy
\be\label{cic}
\text{CIC}<\text{CIC}_\text{min}+\kappa 
\ee
where $s$ is in the region of $0.2$ and $\kappa=5$ yields very solid results. Typically the BIC produces models with few parameters and small errors and generally disfavours large variations in $w(z)$ unless strongly warranted by the data; the AIC on the other hand renders more featured reconstructions, at the expense of larger errors. We find $s=0.2$ works well for the models we present below, balancing AIC and BIC. For alternative data sets, $s$ and the choice of $\kappa=5$ can be adjusted (e.g., $\kappa=10$ is more robust). 

The reconstructed cosmological constant discussed above is shown in Fig~\ref{lambda}.  In this case where $s=0.2$, the reconstruction is good for $z\lesssim1$, but degrades above $z\simeq1.4$. For $s\gtrsim0.8$, however, the fits at high $z$ improve dramatically with the reduced freedom in the fitting functions; this is comparable in complexity to fitting a constant $w$, and so the errors are very tight. This result improves significantly on previous non-parametric reconstructions of $\Lambda$ using SNAP-like data, such as in \cite{Alam:2003sc}. 

\paragraph{Results}

In Fig.~\ref{ws} we show the method in action for two very different types of $w$. One is a standard slow evolution, given by $w=\frac{1}{2}\left\{-1+\tanh\left[3\left(z-\frac{1}{2}\right)\right]\right\}$; the other mimics the effective $w(z)$ one finds in best fit void models of dark energy~\cite{FLSC}, which we model by $w=0.2+1.8\exp\left[-(z-0.4)^2/0.15\right]$. We use $s=0.2$ and $\zeta^\text{fiducial}=\zeta^\text{true}$.
We can see that the reconstruction is impressive, with errors $\sim0.1$. Above $z\sim1$ the errors grow uncontrollably, leading to weak constraints despite the large number of SNIa in this range. This is because constant errors in $\mu$ lead to strongly divergent errors in $w$ in a non-parametric setting. We find that similar fits for other $w(z)$ models which have features on the same scales in $z$-space. We find that if the underlying $w(z)$  contains sharp features such that its derivative is large, then a choice of $s=0.2$ isn't sufficient, and the CIC needs to be weighted more heavily towards AIC (smaller s), which increases the errors. Choosing $s=0.05$ with $N_\text{max}=10$ lets us reconstruct a step-like $w$ with a step of width $\sim0.1$, or a Gaussian bump of about twice that width. Thus, we see that $s$ (combined with the choice of largest $N$, and a choice of $\kappa$) sets the resolving scale for the reconstruction and must be treated as a prior, representing one's intuition of the complexity of $w(z)$. For example, we can reconstruct $w=-1$ to high accuracy by using $s=1$, shown in Fig.~\ref{lambda} (right). This gives errors on $w$ of less than 5\% to $z\sim1$. This situation is analogous to the choice of eigenfunctions which minimise the risk around $\Lambda$ in~\cite{HS}. 

As a final example we consider the $w(z)$ we obtain from the Constitution data set~\cite{hicken}. To do this, we first find the best fit LCDM model, which gives $\zeta=\{0.32, -0.09, .653\}$, and we use this for both the residual $\mu$ and for the model reconstruction. 
\begin{figure}[t]
\begin{center}
\includegraphics[width=\columnwidth]{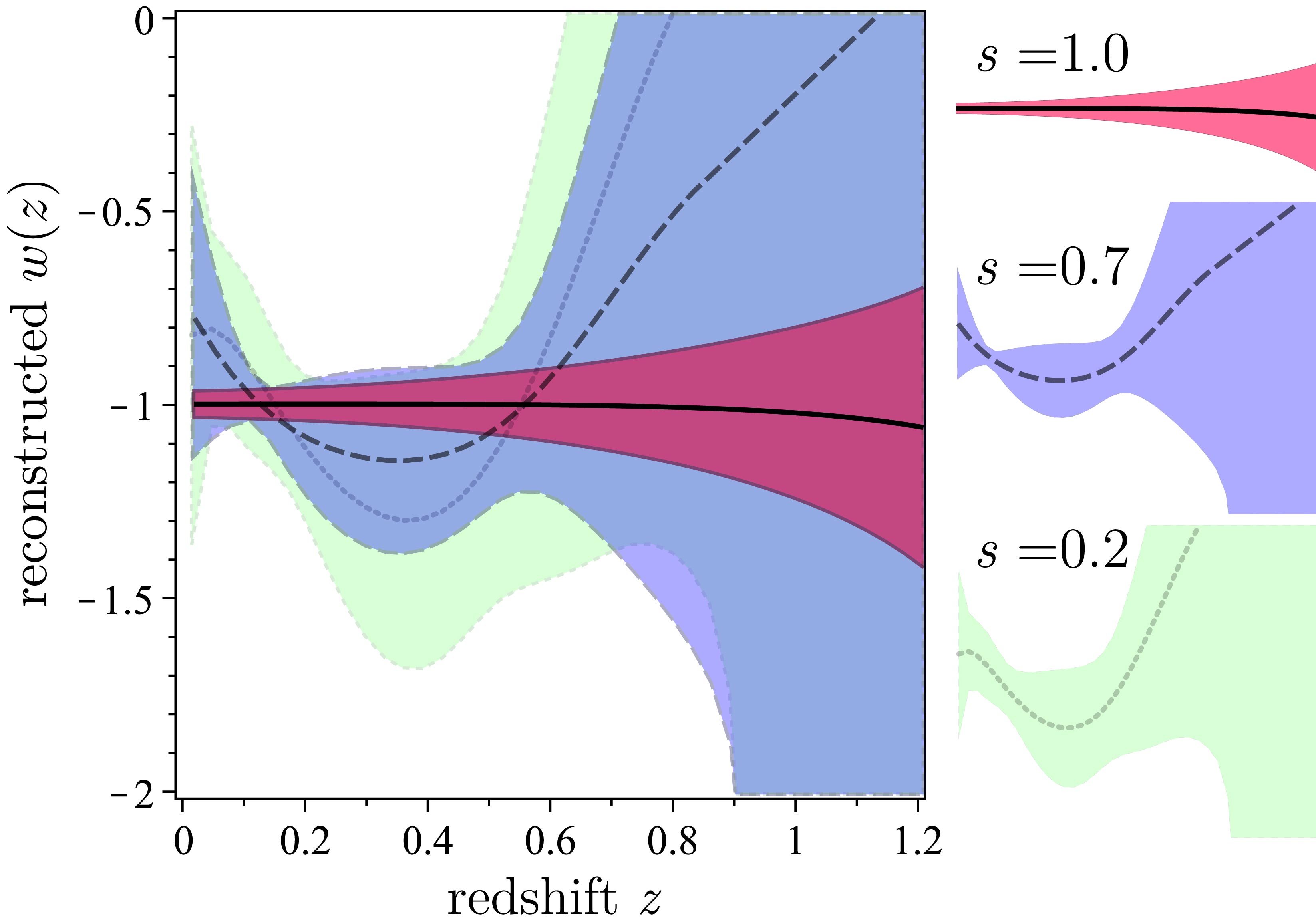}
\caption{Reconstruction of $w$ using the constitution SNIa. }
\label{consrecon}
\end{center}
\end{figure}
In Fig.~\ref{consrecon} we show the constraints on $w(z)$ with different choices of $s$, and using $\kappa=5$. This serves as an illustration of the method and the effect $s$ has, but errors on $\zeta$ have not been folded in for clarity. Using $s=1$ yields tight constraints on a par with those found assuming a constant $w$~\cite{hicken}. This results from using BIC as a selection criterion which yields conservative reconstructions, at the risk of smoothing over real features in the data and underestimating the errors.  
For realistic constraints from a given data set, simulated data of the same standard should be examined first to quantify sensible choices for $s$ and $\kappa$. 

\paragraph{Discussion}
We have presented a novel method to perform a direct reconstruction of $w(z)$, shown to be capable of constraining $w$ to $\sim0.1-0.2$ for $z\lesssim1$ using SNAP-quality data. The method capitalizes on the use of principal component  analysis to find orthogonal bases with which to fit the data, and uses a combination of information criteria to construct a family of reasonably good fitting functions.  Then, by combining the different fitting functions we find the `real' structure present in the data, and smooth over noise-induced features. All fitting is linear, so calculations and errors estimations are easy, with no exploration of parameter space required. Future work will incorporate errors from other parameters, and explore an iterative approach to finding $\mu_\text{resid}$ based on the best fit obtained.

\acknowledgements We thank  Mat Smith for discussions. CC is funded by the NRF (South Africa). CZ is funded by the PIRE grant and the NRF (South Africa).

\end{document}